\documentclass[sigconf,screen]{acmart}
\usepackage[utf8]{inputenc}
\usepackage{enumerate}
\usepackage{amsmath,amssymb,amsfonts}
\usepackage{verbatim}
\usepackage{listings}
\usepackage{hyperref}
\usepackage{csquotes}
\definecolor{gray}{RGB}{96,96,96}
\definecolor{gold}{RGB}{128,128,0}
\definecolor{red}{RGB}{127 0 0}
\definecolor{darkorange}{RGB}{224,64,0}
\definecolor{brown}{RGB}{128,64,0}
\definecolor{navy}{RGB}{0,0,160}
\definecolor{turqoise}{RGB}{0,127,160}
\definecolor{darkgreen}{RGB}{0,127,0}
\definecolor{lila}{RGB}{127,0,160}
\newcommand{\HIDE}[1]{}

\author{John Lång}
\orcid{0000-0002-9394-2464}
\affiliation{%
  \institution{University of Helsinki, Finland; also
  Utrecht University, the Netherlands
  }
}
\email{john.lang@mykolab.com}
\author{I.S.W.B. Prasetya}
\orcid{0000−0002−3421−4635}
\affiliation{%
  \institution{Utrecht University, the Netherlands}
}
\email{s.w.b.prasetya@uu.nl}

\title{Model Checking a C++ Software Framework, a Case Study}

\acmPrice{}
\acmDOI{10.1145/3338906.3340453}
\acmYear{2019}
\copyrightyear{2019}
\acmISBN{978-1-4503-5572-8/19/08}
\acmConference[ESEC/FSE '19]{Proceedings of the 27th ACM Joint European Software Engineering Conference and Symposium on the Foundations of Software Engineering}{August 26--30, 2019}{Tallinn, Estonia}
\acmBooktitle{Proceedings of the 27th ACM Joint European Software Engineering Conference and Symposium on the Foundations of Software Engineering (ESEC/FSE '19), August 26--30, 2019, Tallinn, Estonia}

\begin{document}

    \begin{abstract}
        This paper presents a case study on applying two model checkers,
        {\sc Spin} and {\sc Divine}, to verify key properties of a C++
        software framework, known as ADAPRO, originally developed at
        CERN. {\sc Spin} was used for verifying properties on the
        design level. {\sc Divine} was used for verifying simple test
        applications that interacted with the implementation. Both
        model checkers were found to have their own respective sets of
        pros and cons, but the overall experience was positive. Because
        both model checkers were used in a complementary manner, they
        provided valuable new insights into the framework, which would
        arguably have been hard to gain by traditional testing and
        analysis tools only. Translating the C++ source code into the
        modeling language of the {\sc Spin} model checker helped to
        find flaws in the original design. With {\sc Divine}, defects
        were found in parts of the code base that had already been
        subject to hundreds of hours of unit tests, integration tests,
        and acceptance tests. Most importantly, model checking was
        found to be easy to integrate into the workflow of the software
        project and bring added value, not only as verification, but
        also validation methodology. Therefore, using model checking
        for developing library-level code seems realistic and worth the
        effort.
    \end{abstract}

\ccsdesc[500]{Software and its engineering~Formal software verification}

\keywords{model checking concurrent C++, verification concurrent C++, model checking C++ case study}

    \maketitle

	\section{Introduction}
	    ADAPRO stands for \emph{ALICE Data Point Processing Framework}.
        It is an open source C++ 14 software framework\footnote{available online at \hyperref[https://gitlab.com/jllang/adapro]{https://gitlab.com/jllang/adapro}}, consisting of
        about 6000 lines of source code\footnote{as measured in 25
        March 2019, using David A. Wheeler's Sloccount utility,
        available online at
        \hyperref[https://dwheeler.com/sloccount/]{https://dwheeler.com/sloccount/}}.
        It is meant for building configurable, remote-controllable,
        multi-threaded daemon applications. ADAPRO was originally
        conceived as a collection of common routines used for
        implementing the \emph{ALICE Data Point Service}
        (ADAPOS)\cite{Laang} software architecture (as part of the
        ALICE RUN3 upgrade\cite{Abelev}). ALICE stands for
        \emph{A Large Ion Collider Experiment}.
        ALICE is one of the four major experiments at the
        \emph{Large Hadron Collider} (LHC) of the \emph{European
        Organisation for Nuclear Research} (CERN).

        The highly concurrent nature of ADAPRO makes its verification
        challenging by conventional means. Although the framework and
        its applications\cite{Laang} have been subject to hundreds of
        hours of unit tests, integration tests, and acceptance tests,
        ADAPRO is too complex for its all behaviours to be anticipated
        by tests. Therefore, it is also hard to say how adequate these
        tests actually were in covering the software's concurrent
        behavior. Even though ADAPRO has its roots in a specific use
        case, it has evolved into a reusable tool. Some ADAPRO
        applications may be expected to be able to run autonomously for
        months without human intervention.

        All of these facts reinforce the importance of a formal
        verification project, because even rarely occurring defects may
        cause costly damage, as noted by Gerard J. Holzmann, the author
        of the {\sc Spin} model checker \cite{Holzmann2001}. A study by
        John Fitzgerald et al. confirms the impact of formal
        verification on software quality \cite{Fitzgerald2013}.

        To address this challenge we set up a project to explore the
        feasibility of applying
        {\em model checking}\cite{Jhala,Rozier,BaierKatoen} to
        thoroughly verify ADAPRO's critical properties. We chose model
        checking, because it naturally fits with the \emph{Finite
        State Machine} (FSM) paradigm used by the framework, which
        will be discussed later. Furthermore, we decided to use model
        checking on two levels of abstraction. This decision proved
        to be a good idea, since using two different model checkers in a
        complementary manner helped to find different kinds of issues
        faster.

        On the higher level of abstraction, we wanted to verify that the
        very design of ADAPRO itself is correct, so we constructed a
        new model for it. After experimenting with the
        \emph{NuSMV}\cite{CAV02}, \emph{TLA{+}}\cite{Lamport2002}, and
        {\sc Spin}\cite{SPIN} model checkers, we decided to choose
        {\sc Spin} for this purpose. For verifying ADAPRO's actual
        implementation, we performed software model checking. The model
        checker {\sc Divine}\cite{BBK+17} was chosen for this purpose.

        \subsubsection*{Contribution}
        Formal verification turned out to reveal important issues not
        previously found by testing, even though it wasn't possible to
        obtain exhaustive results. Our findings necessitated changes in
        the design and implementation of ADAPRO. The changes will be
        part of the foundation of the next major version of ADAPRO
        (v5.0.0\footnote{we discuss the version available at
        \hyperref[https://doi.org/10.5281/zenodo.3258225]{https://doi.org/10.5281/zenodo.3258225}}). In this
        paper we will share our experience, discoveries, and lessons
        gained from our verification project.

        \subsubsection*{Paper structure}
        We first give a brief overview on related work in Section 2.
        Section 3 offers an overview of the ADAPRO framework. Section 4
        gives the necessary definitions, assumptions and properties for
        design-level verification. In Section 5, we discuss our
        findings on this part of the project. Section 6 describes the
        effort for verifying the C++ implementation and its results. In
        Section 7, we discuss our experiences and lessons learned from
        using the two model checkers. We end the article with
        conclusion, future prospects, and acknowledgements.

    \section{Related Work}

        There are many case studies on using model checking to verify
        the correctness of production software; too many to list here
        comprehensively. Some examples include
        \cite{Adiego, Chandra, Chen, delaCamara, Gan, Holzmann99, Hwong, Wing}.
        Fitzgerald et al. have written a survey on
        many industrial use cases \cite{Fitzgerald2013}.

        ADAPRO is similar to the \emph{SMI++}\cite{SMI++} system in its
        reliance on an FSM model. However, ADAPRO is more restricted,
        since it doesn't have a \emph{Domain Specific Language} (DSL)
        and its FSM model is rigid in the sense that the user can't
        define new states or commands. Instead of synthesizing
        distributed control systems, ADAPRO is focused on assembling
        threads into a remote controllable concurrent application.
        ADAPRO applications can interact with SMI++ based systems using
        the \emph{Distributed Information Management}
	    (DIM)\cite{GASPAR2001102} protocol as its communications layer,
	    which has been demonstrated by the ADAPOS Manager application.

	    All LHC experiments use the SMI++
	    system for their control systems. Formal verification has been
	    performed on the \emph{Compact Muon Solenoid} (CMS) control
	    system \cite{Hwong}, which demonstrates the feasibility
	    of building and analysing control systems with tens of thousands
	    of nodes, based on hierarchies of FSMs. ADAPRO also follows a
	    similar approach, though it features a simple tree with just
	    root and a number of leafs.

	    Compared to most of the papers mentioned in this section,
	    our approach was more lightweight in that we didn't use
	    automated model extraction or translation tools. Similarly to
	    \cite{Wing} and \cite{Chen}, we wanted to explore the correct
	    level of abstraction for finding the most relevant aspects of
	    the algorithms used in ADAPRO, by building the design-level
	    model by hand. Due to the complexity of ADAPRO,
	    full formal software model checking was not computationally
	    feasible. Instead, we used {\sc Divine} as a high coverage bug
	    hunting tool. Our lightweight approach seemed to suit the needs
	    of this project well, and we believe it to be realistically
	    reproducible in other similar projects.

    \section{ADAPRO}\label{sec.adapro}
\begin{figure}
            \includegraphics[scale=0.75]{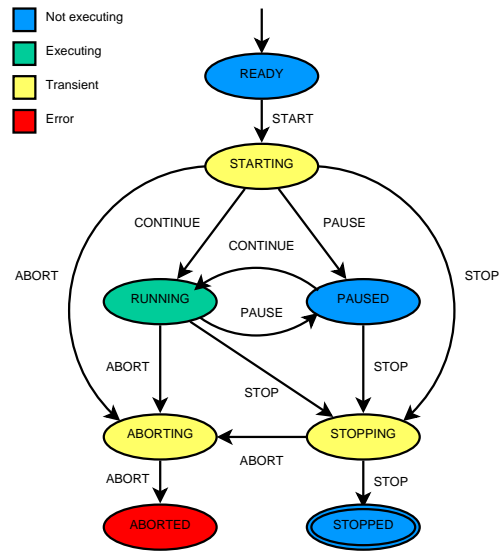}
            \caption{Thread state transition diagram}
            \label{fig:stateDiagram}
        \end{figure}

        The basic actor in ADAPRO is the abstract \emph{Thread} class,
        which follows an FSM approach. The domain logic of an
        application is meant to be implemented as virtual methods
        and/or callbacks, called \emph{user-defined code}, provided to
        the framework through specialized Thread instances. The role of
        the framework is to manage Threads. Figure
        \ref{fig:stateDiagram} shows the state transition diagram of
        the Thread FSM.

        A Thread moves from one state to another upon {\em commands} as
        depicted in Figure \ref{fig:stateDiagram}. Should the Thread
        ever encounter an undefined state/command combination, it is
        specified to take no action (other than printing a warning
        message) in such situation. To prevent this from happening, the
        FSM is not directly exposed to the user. Instead, an
        object-oriented approach with safe accessor methods is used.
        The methods for changing the command of a Thread are called
        \emph{trigger methods}, and they have \emph{synchronous}
        (i.e. blocking) and \emph{asynchronous} (non-blocking)
        variants. More information on the
        \emph{Application Programmable Interface} (API) is available on GitLab\footnote{see
        \hyperref[https://gitlab.com/jllang/adapro/-/jobs/artifacts/5.0.0/download?job=manual]{https://gitlab.com/jllang/adapro/-/jobs/artifacts/5.0.0/download?job=manual}
        for the manual and
        \hyperref[https://jllang.gitlab.io/adapro/]{https://jllang.gitlab.io/adapro/} for the API documentation}.

        During a state transition, the Thread\footnote{throughout this
        article, we use capital initial letter `T' to distinguish the
        ADAPRO class Thread from the general concept of a thread}
        invokes a \emph{transition callback}, a user-defined function
        that takes the target state as its argument. The transition
        callback is not allowed to throw exceptions or interfere with
        the framework in any other way. If the Thread enters the state
        {\tt STARTING} or {\tt STOPPING}, it invokes a corresponding
        virtual method, implemented by the user. The Thread keeps
        invoking another virtual method in a loop for as long as it
        stays in the state {\tt RUNNING}. The three virtual methods
        associated with these states ({\tt prepare}, {\tt finish}, and
        {\tt execute} respectively) are allowed to fail by throwing
        exceptions. If the Thread base class catches an exception
        thrown by the inheriting class, it moves to the state
        {\tt ABORTING}, prints an error message, moves to the state
        {\tt ABORTED}, and halts.

        The states whose names end with ``{\tt ING}'' in Figure
        \ref{fig:stateDiagram}
        are the ones, during which the backend thread is performing
        computation. A Thread that is in one of the states {\tt READY},
        {\tt STOPPED}, or {\tt ABORTED} must neither posess any
        dynamically allocated objects or resources, nor hold any locks,
        so that it can be safely deleted. A Thread in state
        {\tt PAUSED} is considered to be temporarily suspended and
        capable of moving back to the state {\tt RUNNING} or proceeding
        to the {\tt STOPPING} state on short notice.

        Thread uses the C++ standard library thread class as its
        backend. Some advanced functionality also involves the PThreads
        and Linux APIs through native handles. POSIX or Linux specific
        parts of the framework are not critical and they’re guarded
        with macros to ensure portability. These parts were left
        outside of the scope of this verification project.

\begin{figure}
            \includegraphics[scale=0.7]{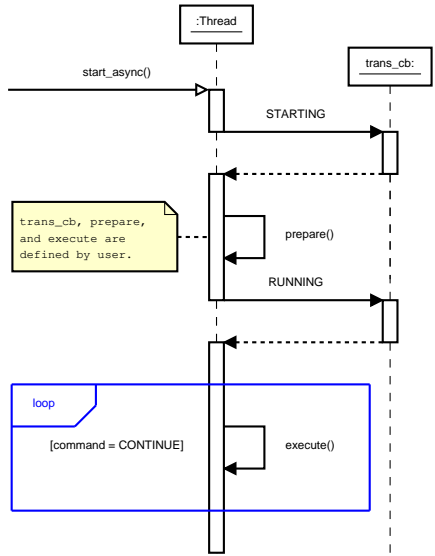}
            \caption{Asynchronous Thread startup}
            \label{fig:asyncStartup}
        \end{figure}

        The interpretation of the states and transitions is up to the
        implementing class, since Thread doesn't define (i.e. it only
        \emph{declares}) the virtual methods or the transition
        callback. The intended use case of the Thread class is running
        a repetitive task in background, in the way how services or
        daemons work in process level. For a batch job, a standard
        library thread is probably more appropriate choice. Figure
        \ref{fig:asyncStartup} is a UML sequence diagram showing the
        asynchronous startup sequence of a Thread as an example on the
        interactions between the owner of a Thread, the backend of the
        Thread, and the transition callback.

    \subsection{Session and Supervisor}
        An ADAPRO application consists of a special Thread known as the
        \emph{Supervisor} and one or more other Threads, known as
        workers. The user mustn't manage Supervisor directly, but
        instead utilize the static methods of a class called
        {\em Session}, which follows the \emph{singleton} design
        pattern and runs on the main thread. Session is responsible for
        framework startup and shutdown sequences. For technical
        reasons, the lifetime of a Session consists of separate
        initialization and runtime phases.

\begin{figure}
            \includegraphics[scale=0.75]{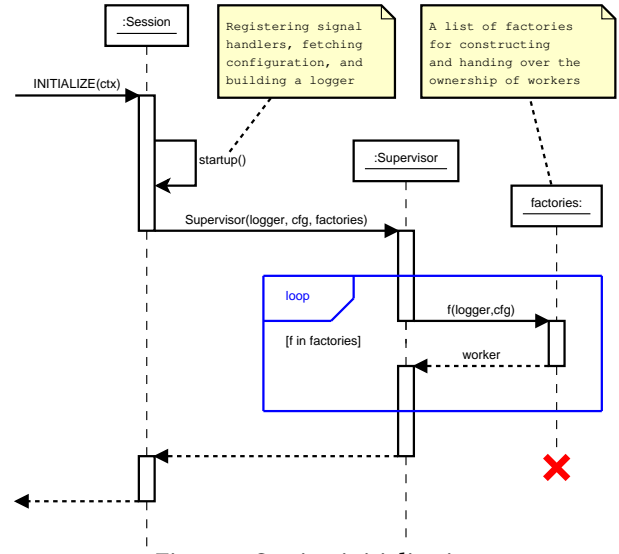}
            \vspace{-4mm}
            \caption{Session initialization}
            \label{fig:initialization}
        \end{figure}

        During the initialization phase, Session registers signal
        handlers first. Then it constructs the configuration using
        default values, file, and/or command-line arguments,
        initializes a logger. After that, Session possibly performs
        certain interactions with the operating system. At the end of
        the initialization phase, Session constructs a Supervisor,
        handing over references to the logger, configuration, and
        user-defined worker factories. Supervisor then takes care of
        constructing the workers by applying the factories to the
        logger and configuration. Figure \ref{fig:initialization} gives
        a simplified overview on Session initialization.

\begin{figure}
            \includegraphics[scale=0.75]{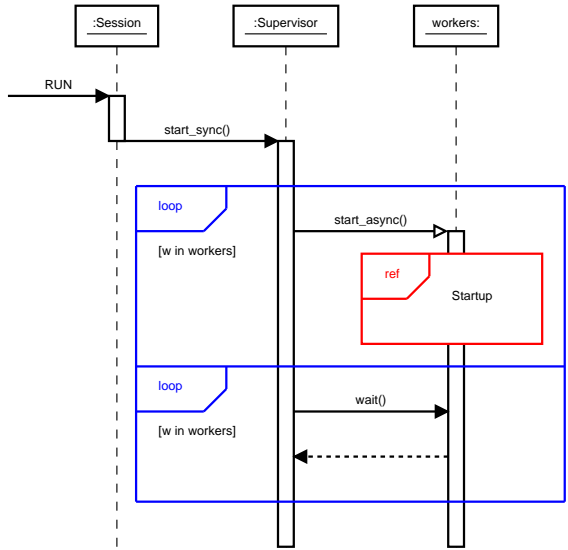}
            \caption{Command propagation}
            \label{fig:runtime}
        \end{figure}

        The state of Supervisor represents the overall
        state of the application in the sense that Supervisor is the
        first Thread to start and the last Thread to stop. When a
        signal or command arrives from an external process, Supervisor
        propagates the appropriate FSM command to the workers. Figure
        \ref{fig:runtime} presents propagation of the {\tt START}
        command as an example. The sequences for propagating
        {\tt PAUSE}, {\tt RESUME}, and {\tt STOP} commands are similar.

        The runtime phase of Session starts with Session sending
        the {\tt START} command to Supervisor. After that, Session
        remains passive until the Supervisor halts or a signal handler
        or the global exception handler is activated. When the runtime
        phase ends, Session returns a one-byte status code, which is a
        bitmask of eight different flags representing certain common
        error categories.

        Supervisor starts workers asynchronously, after which it blocks
        until all workers have ended their startup sequences. When the
        startup sequence ends, Supervisor and all workers have
        moved into one of the six states below {\tt STARTING} shown in
        Figure \ref{fig:stateDiagram}. In addition to propagating
        an external {\tt STOP} command, Supervisor also propagates it
        if one or more workers have aborted or if all of the workers
        have stopped.

    \section{Design-Level Verification}
        We first wanted to verify that the design of ADAPRO is correct.
        To this end, a set of key correctness properties called
        \emph{the Theory of ADAPRO} was identified and formalized in
        \emph{Linear Temporal Logic} (LTL)\cite{Pnueli}, presented in
        Section \ref{subsec.theory} below. Section
        \ref{subsec.model.adapro} discusses the construction a
        model capturing ADAPRO's logic. The model was written in
        \textsc{Promela}\cite{SPIN}. Using the model checker Spin
        \cite{BenAri,SPIN}, we verified that this model satisfies the
        Theory of ADAPRO. Section \ref{findings} discusses the findings.

        \subsection{Preconditions}
		The ADAPRO framework relies on a programming contract between
		the framework developers and the application developers. There
		are necessary \emph{preconditions}/assumptions on the
		environment and user-defined code, that ADAPRO has to take as
		granted. The full list of preconditions is given below for
		completeness:
		\begin{enumerate}
		    \item
		        There will be no sabotage or force majeur incidents
		        of any kind whatsoever.
		    \item
		        The C++ compiler, standard library and runtime
		        system, the operating system and the hardware
		        platform function correctly with respect to their
		        specifications.
		    \item
		        The operating system scheduler satisfies weak fairness,
		        i.e. the property that every native thread that is
		        eventually always executable, will be run on a
		        processor unboundedly many times.
		    \item
		        Allocating memory for new objects always succeeds,
		        i.e. the machine will not run out of (virtual)
		        memory.
		    \item
		        User-defined code always terminates.
		    \item
		        Unless explicitly permitted to do so, by ADAPRO
		        manual or the API documentation, user-defined code
		        will not modify or delete any object owned by the
		        framework.
		    \item
		        User-defined code never
		        \begin{itemize}
		            \item
		                calls {\tt std::abort};
		            \item
		                raises a signal that can't be handled;
		            \item
		                causes deadlocks or livelocks;
		            \item
		                uses Thread trigger methods inappropriately; or
		            \item
		                triggers a known issue in the framework.
		        \end{itemize}
		    \item
		        User-defined code conforms to the well-known
		        \emph{Resource Aqcuisition is Initialization} (RAII)
		        principle, so that all memory and resources
		        acquired by user-defined code will be automatically
		        released when ownership expires (e.g. a destructor
		        of a user-defined object, holding a resource, is
		        invoked).
		    \item
		        In DIM server mode or daemon mode, the
		        user-defined code does not directly interact with
		        DIM or Systemd libraries respectively.
	    \end{enumerate}

	    The appropriateness of the use of trigger methods is best
	    explained by enumerating the acceptable scenarios. We do this
	    in Sections \ref{verification} and \ref{divineModels} for the
	    design-level and the implementation-level models respectively.
    \subsection{The Theory of ADAPRO}
        \label{subsec.theory}
        Let us first define the following notation:

		\newcommand{\pnull}         {\operatorname{null}}
		\newcommand{\ready}         {\operatorname{ready}}
		\newcommand{\starting}      {\operatorname{starting}}
		\newcommand{\running}       {\operatorname{running}}
		\newcommand{\paused}        {\operatorname{paused}}
		\newcommand{\stopping}      {\operatorname{stopping}}
		\newcommand{\stopped}       {\operatorname{stopped}}
		\newcommand{\aborting}      {\operatorname{aborting}}
		\newcommand{\aborted}       {\operatorname{aborted}}
		\newcommand{\halting}       {\operatorname{halting}}
		\newcommand{\halted}        {\operatorname{halted}}
		\newcommand{\executable}    {\operatorname{executable}}
		\newcommand{\executing}     {\operatorname{executing}}
		\newcommand{\opU}           {\mathbf{U}}
		\newcommand{\opW}           {\mathbf{W}}

        \begin{enumerate}[(i)]
            \item
                Denote Supervisor as $s$;
            \item
                The set of all Threads, including Supervisor, as $T$;
            \item
                The set of Workers, i.e. $T \setminus \{s\}$ as $V$;
            \item
                $\pnull(x)$ as the predicate ``Thread $x$ doesn't
                exist'';
            \item
                For each state $q$ the predicate $q(x)$ expressing that
                ``Thread $x$ is in state $q$'' (e.g. $\ready(x)$ denotes
                ``Thread $x$ is in state {\tt READY}'');
            \item
                $\halting(x)$ as $\stopping(x) \vee \aborting(x)$;
            \item
                $\halted(x)$ as $\stopped(x) \vee \aborted(x)$;
            \item
                $\executable(x)$ as ``Thread $x$ is in the state
                {\tt RUNNING} with its command set to {\tt CONTINUE}'';
                and
            \item
                $\executing(x)$ as ``Thread $x$ is carrying out its
                action associated with the state {\tt RUNNING}''.
        \end{enumerate}

        The following are the key correctness properties of ADAPRO,
        that represent the \emph{postconditions} promised as a part of
        the programming contract:

        \begin{enumerate}
            \item \label{spec.ready}
                $\forall_{t \in T}[\pnull(t) \opU
                    (\ready(t) \opW \starting(t))]$;
            \item \label{spec.starting}
                $\forall_{t \in T} \square [\starting(t) \rightarrow \\
                (\starting(t) \opU (\running(t) \vee \paused(t) \vee \halting(t)))]$;
            \item
                $\forall_{t \in T} \square [\running(t) \rightarrow \\
                    (\running(t) \opW (\paused(t) \vee \halting(t)))]$;
            \item
                $\forall_{t \in T} \square [\paused(t) \rightarrow \\
                    (\paused(t) \opW (\running(t) \vee \stopping(t)))]$;
            \item \label{spec.stopping}
                $\forall_{t \in T} \square [\stopping(t) \rightarrow \\
                    (\stopping(t) \opU (\stopped(t) \vee \aborting(t)))]$;
            \item
                $\forall_{t \in T} \square [\stopped(t) \rightarrow
                    \square \stopped(t))]$;
            \item \label{spec.aborting.leadsto.aborted}
                $\forall_{t \in T} \square [\aborting(t) \rightarrow
                    (\aborting(t) \opU \aborted(t))]$;
            \item
                $\square (\forall_{t \in T} [\aborted(t) \rightarrow
                    \square \aborted(t))])$;
            \item
                $\square
                    (\ready(s) \rightarrow \forall_{v \in V}[\pnull(v)])$;
            \item
                $\square
                    (\halted(s) \rightarrow \forall_{v \in V}[\pnull(v)])$;
            \item
                $\square
                    (\exists_{v \in V}[\ready(v)] \rightarrow \starting(s))$;
            \item
                $\square (\forall_{v \in V}[\stopped(v)] \rightarrow
                    (\paused(s) \vee \stopped(s)))$;
            \item
                $\square (\exists_{v \in V}[\aborted(v)] \rightarrow
                    (\paused(s) \vee \stopped(s)))$;
            \item
                $\square (\halting(s) \rightarrow
                    \lozenge \forall_{v \in V}[\halted(v)])$;
            \item
                $\neg \lozenge \aborted(s)$;
            \item
                $\forall_{t \in T}[\lozenge \square \executable(t)
                    \rightarrow \square \lozenge \executing(t)]$; and
            \item
                $\forall_{t \in T}[\square \lozenge (\executable(t)
                    \rightarrow \executing(t))]
                    \opW \halting(s)$.
        \end{enumerate}
        Formulae 1 -- 8 capture the next-state relation induced by the
        FSM (see Figure \ref{fig:stateDiagram}). Notice that some of
        them (\ref{spec.ready}, \ref{spec.starting},
        \ref{spec.stopping}, \ref{spec.aborting.leadsto.aborted}), use
        strong until ($\opU$), to express that the corresponding
        transitions are inevitable. Others use weak until to express
        that the transitions are not required to be taken.
        
        Formulae 9 -- 15 express additional safety properties
        that every ADAPRO session is expected to satisfy, e.g. 9 and 10
        state that no worker should exist while the Supervisor is still
        in the {\tt READY} state, or after it halts, whereas 15 says
        that the Supervisor should never abort. Formula 16 expresses
        weak fairness. Finally, formula 17 is a liveness property that
        asserts that under the right conditions, all Threads get to
        execute their main task.

     \subsection{Modeling Strategy} \label{verification}
        Since a model checker cannot verify a generic system with
        unbounded number of Thread instances, we model a system
        representing an ADAPRO application with one Supervisor and two
        workers. The rationale for using two workers is that it allows
        the distinction between universally and existentially quantified
        statements about workers. Having two workers also allows the
        model checker to expose potentially inconsistent worker state
        combinations. We suspect that having three or more workers
        would only make the model larger without introducing
        essentially new species of errors. As noted in the beginning of
        Section \ref{findings}, state space size was a practical reason
        forcing us to limit our scope to two workers.

        The workers are treated as black boxes that only have their FSM
        interface and do not communicate with each other or share
        resources. A worker may initiate the following state
        transitions, simulating unhandled exceptions and the use of the
        worker's own trigger methods\footnote{in C++, a Thread must
        always call its own trigger methods asynchronously; otherwise
        it will end up in a deadlock waiting for itself to finnish
        executing its own command. This issue was known and documented
        well before beginning the verification project.}:
        \begin{itemize}
            \item
                During {\tt prepare}, the worker chooses
                non-deterministically between pausing, stopping,
                aborting, and doing nothing.
            \item
                During {\tt execute}, the worker chooses
                non-deterministically between stopping, aborting, and
                doing nothing.
            \item
                During {\tt finish}, the worker chooses
                non-deterministically between aborting and doing
                nothing.
        \end{itemize}

        Supervisor was modeled quite faithfully with respect to its C++
        implementation. In the model, Supervisor doesn't spontaneously
        send commands to workers. It delegates commands received from
        the init process, which represents the environment, and reacts
        if both workers stop or at least one of them aborts.

    \subsection{Building the \textsc{Promela} Model}
        \label{subsec.model.adapro}
        Since ADAPRO initially did not have a formal design, we
        reconstructed manually its design/logic from its C++
        implementation in the modeling language
        \textsc{Promela}\cite{SPIN}. We focused on the logic that
        implements the FSM in Figure \ref{fig:stateDiagram}.

        To model the behavior of a C++ program inevitably involves
        figuring out how to map C++ programming patterns to
        \textsc{Promela}. For example, the state of an ADAPRO Thread is
        stored in a private field\footnote{or \emph{data member} in C++
        terminology} of type {\tt std::atomic<State>}, where the
        template argument is an enumerated type consisting of the eight
        different states indicated in Figure \ref{fig:stateDiagram}.
        This field is only accessed through the accessors
        {\tt get\_state} and {\tt set\_state}.

        In a concurrent C++ application, naïve reading and writing of
        shared memory does not work as one might expect. Threads
        executing on CPU cores may use the cores' local cache memory.
        It’s not possible to control caching directly using C++
        language constructs. Modifications made by one thread might get
        stuck in the local cache of a CPU core and never become visible
        to other threads.

        Instruction reordering might also wreak havoc by causing
        modifications to show in wrong order. Instead of providing its
        own solutions, ADAPRO relies on C++ standard library
        synchronization primitives. In particular, reading and writing
        a shared state involves a memory barrier to ensure that the
        changes will be visible to all parties as intended:

        \begin{lstlisting}
enum State { READY,    STARTING, RUNNING,  PAUSED,
             STOPPING, STOPPED,  ABORTING, ABORTED };
enum Command { START, STOP, PAUSE, CONTINUE, ABORT };
class Thread {
    std::atomic<State> state;
    std::atomic<Command> command;
    std::mutex m;
    void set_state(const State s) noexcept {
        state.store(s, std::memory_order_release); }
public:
    State get_state() const noexcept {
        return state.load(std::memory_order_consume); } }; \end{lstlisting}

        In contrast, \textsc{Promela} offers high level
        {\em atomicity}: a single assignment is always atomic,
        and it is possible to declare a block of statements to be
        atomic. \textsc{Promela} also guarantees {\em sequential
        consistency}, defined by Leslie Lamport\cite{Lamport79} as
        follows: \blockquote{
        ``[T]he result of any execution is the same as if the operations
        of all the processors were executed in some sequential order,
        and the operations of each individual processor appear in this
        sequence in the order specified by its program.''}

        This implies that all (atomic) changes to variables are
        guaranteed to be visible to all processes. With these
        properties, inter-thread communications in ADAPRO become easy
        to model in \textsc{Promela}, hence allowing us to focus on
        modeling its algorithms. The ADAPRO functionality
        mentioned can be modeled succinctly as as shown below. We assume
        there are $N$ threads; their states are stored in a global
        array called $\tt states$:

        \begin{lstlisting}
mtype = {READY,STARTING,RUNNING,PAUSED,STOPPING,STOPPED,
         ABORTING,ABORTED};
mtype states[N]; /*the states of the N threads*/
#define get_state(i) states[i]
#define set_state(i,s) states[i] = s \end{lstlisting}

       In C++, {\tt std::mutex} and {\tt std::condition\_variable} are
       often used to implement inter-thread communication. In
       \textsc{Promela} we can abstract this away. For example, if a
       process $P$ wants to wait until a certain predicate $C$ becomes
       true, we can simply write $C$ as a guarding expression, hence
       resulting in a cleaner model. The code example below presents a
       simplified Thread model that utilizes the comparisons of the
       command and state values as guards (e.g.
       $\tt get\_command(k) == START$):

\begin{lstlisting}
proctype Thread(byte k) { /* a Thread with id k */
  /* Startup: */
  get_command(k) == START;
  set_state(k,STARTING);
  set_command(k,CONTINUE);
  prepare(k); /* the action of the state STARTING */
  /* Execution of user code: */
  do
  ::get_command(k) == CONTINUE ->
      if
      :: get_state(k)==STARTING-> set_state(k,RUNNING)
      :: get_state(k)==PAUSED  -> set_state(k,RUNNING)
      :: else -> skip
      fi
      execute(k)
  ::get_command(k) == PAUSE  ->
      if
      ::get_state(k)==PAUSED -> get_command(k)!=PAUSED
      ::else ->
          if
          ::get_state(k)==STARTING||get_state(k)==RUNNING
                 -> set_state(k,PAUSED)
          ::else -> skip
          fi
      fi
  ::get_command(k)==STOP||get_command(k)==ABORT -> break
  od
  /* Shutdown: */
  if
  ::get_command(k)==STOP -> /* ... */
  ::get_command(k)==ABORT-> /* ... */
  fi }\end{lstlisting}

        Notice that in the case where the command is either {\tt STOP}
        or {\tt ABORT}, {\tt get\_command(i)} appears twice. Because
        it's a macro expanding to {\tt commands[i]}, it doesn't require
        a new variable. Both comparisons are part of the same atomic
        \textsc{Promela} transaction. Using a temporary variable in a
        situation like this might increase the state space size of the
        model and even break temporal properties as the level of
        atomicity would change. During its execution, a Thread will
        keep looping indefinitely and performing the appropriate action
        as long as its command is not {\tt STOP} or {\tt ABORT}. If the
        command changes to one of these, the Thread exits the loop and
        begins its shutdown sequence.

        The full model is nearly 800 lines long (including comments)
        and can be found at\footnote{\hyperref[https://gitlab.com/jllang/adapro/tree/5.0.0-RC3/models/promela]{https://gitlab.com/jllang/adapro/tree/5.0.0-RC3/models/promela}}.
        The model maintains a quite fine grained atomicity
        in order to maximise interleaving possibilities that will be
        checked. As the tradeoff, model checking a fine grained model
        can be expected to consume more resources.

    \section{Verification Results with Spin}
        \label{findings}

        Verifying the theory of ADAPRO successfully with a Supervisor
        and two workers \textsc{Spin} takes less than three gigabytes
        of memory (without need for compressing states), and a couple
        of minutes with a high-end laptop. For three workers, the
        verification takes {\sc Spin} ran out of the 17 GB of memory
        available to it, after nearly two hours. The verifier did
        manage to verify properties (1) -- (15) for three workers
        though, thanks to state compression option {\tt -DCOLLAPSE}
        used for compilation.

        Writing the model itself exposed a number of design flaws and
        bugs in the C++ code. Additional defects were found during model
        checking because the \textsc{Promela} model was accurate enough
        to replicate them. This section discusses two of the issues
        that had managed to remain undetected during testing.

    \subsection{Revoked ABORT Defect}
        \label{revokedAbort}
        As mentioned in Section \ref{sec.adapro}, a Thread has to move
        to the state {\tt ABORTING} if it encounters an unhandled
        exception. It is important that the {\tt ABORT} command cannot
        be revoked. This requirement is captured by specification
        (\ref{spec.aborting.leadsto.aborted}) in the ADAPRO Theory
        (Section \ref{subsec.theory}). It turned out that this property
        was violated.

        After startup, a Thread runs one loop as shown in the previous
        listing. In C++, the algorithm looks like the following:
        \begin{lstlisting}
void Thread::run() noexcept {
    try {
        /* Startup */
        bool shutdown{false};
        while (!shutdown) {
            switch (get_command()) {
                case CONTINUE: /* ... */ break;
                case PAUSE:    /* ... */ break;
                default: shutdown = true;
            }
        }
        /* Shutdown */
    }
    catch (const std::ios_base::failure& e) {HANDLE(e);}
    catch (const std::system_error& e) {HANDLE(e);}
    /* ... */
    catch (const std::exception& e) {HANDLE(e);} } \end{lstlisting}

        The macro {\tt HANDLE} takes care of printing an
        error message and initiating the transition to state
        {\tt ABORTED} via {\tt ABORT}. Multiple {\tt catch} blocks with
        the same code are needed to avoid object slicing, which would
        cause the method {\tt std::exception::what} to return just
        {\tt "std::exception"} (on Linux systems at least), which is
        not a very helpful error message.

        It is possible that user-defined code throws an exception,
        which is caught by one of the {\tt catch} blocks. The method
        {\tt handle\_exception} then sets the command to {\tt ABORT}.
        At this point, it may happen that a {\tt PAUSE} command arrives
        from an external source, and that Supervisor propagates the
        command {\tt PAUSE} to all workers. Now the worker that was
        about to abort is told to go to the state {\tt PAUSED} instead.
        This violates the FSM constraints (see Fig.
        \ref{fig:stateDiagram}), as {\tt STOP} and {\tt ABORT} should
        never get overridden by commands of lower priority.

        To make sure that {\tt set\_state} can never violate the FSM
        mechanism, and that the {\tt ABORT} command is irrevokable, it
        was necessary to redefine {\tt set\_command} (which was
        previously defined in a fashion similar to {\tt set\_state}) in
        \textsc{Promela} as follows:

        \begin{lstlisting}
#define set_command(i, c)                               \
    atomic {                                            \
        if                                              \
        ::  c == START && commands[i] == CONTINUE ->    \
                commands[i] = START                     \
        /* ... */                                       \
        ::  c == STOP && (commands[i] == CONTINUE ||    \
                          commands[i] == PAUSE)   ->    \
                commands[i] = STOP                      \
        ::  c == ABORT ->  commands[i] = ABORT          \
        fi } \end{lstlisting}

        The corresponding update in the C++ code required changing the
        locking scheme. It took a few attempts to arrive at a correct
        solution. The {\sc Divine} model checker was found useful
        during this development process, as it discovered a flaw in one
        of the attempted solutions. The final correct version is given
        below:
        
\lstset{
    classoffset=1,
    otherkeywords={atomic,mutex,condition_variable,unique_lock,
        lock_guard,uint8_t,State,Command,Thread,byte,failure,
        system_error,exception},
    morekeywords={atomic,mutex,condition_variable,unique_lock,
        lock_guard,uint8_t,State,Command,Thread,byte,failure,
        system_error,exception},
    keywordstyle=\color{navy},
    classoffset=5,
    otherkeywords={wait_for_state,_mask,state_in},
    morekeywords={wait_for_state,_mask,state_in},
    keywordstyle=\color{black}
    }
        \begin{lstlisting}
void Thread::set_command(const Command c) noexcept {
    bool success{false};
    std::lock_guard<std::mutex> lock{m};
    switch (c) {
        case START:
            success = get_state() == READY
                    && get_command() == CONTINUE;
            break;
        case PAUSE:
            success = state_in(STARTING | RUNNING)
                    && get_command() == CONTINUE;
            break;
        /* ... */
        case ABORT:
            success=state_in(STARTING|RUNNING|STOPPING);
    }
    if (success) {
        command.store(c, std::memory_order_release);
    } } \end{lstlisting}
        The method {\tt run} had to be changed as well, to hold a lock
        on the mutex when performing state transitions. The lock has to
        be released before, and reacquired after, executing user-defined
        virtual methods, though. The reason is that the user-defined
        virtual methods are allowed to use certain trigger methods of
        the executing Thread (see Section \ref{verification}).

    \subsection{A Synchronisation Defect}
        \label{synchronisationDefect}
        {\sc Spin} was also to able to re-discover a known issue.
        There was an overlooked design flaw in the C++ code for a
        synchronisation method named {\tt wait\_for\_state}, part of
        the API. It had escaped CPPUnit tests of ADAPRO that existed at
        the time, which had around 90\% overall line coverage.

        The method {\tt wait\_for\_state} is supposed to block until the
        Thread has moved to a state greater than or equal to a given
        target state. In terms of executions, some states come after
        others, so they are greater than their predecessors. As ADAPRO
        leaves scheduling to the operating system, it cannot guarantee
        that {\tt get\_state} will be executed exactly at the right
        moment, and not after the Thread has already moved to a later
        state, hence the need for also accepting states greater than the
        target state.

        Two synchronization primitives have to be added to the Thread
        class in order to implement blocking behaviour. The C++ code
        snippet below shows the erroneous
        {\tt wait\_for\_state} implementation. It relies on the
        integer representation underlying the enumerated type
        {\tt State} for the comparison {\tt get\_state() < s}:
        \begin{lstlisting}
class Thread {
    std::atomic<State> state;
    std::mutex m;
    std::condition_variable cv;
    void set_state(const State s) noexcept {
        state.store(s, std::memory_order_release);
        cv.notify_all(); // This new line was needed
    }
public:
    State get_state() const noexcept; // Same as before
    void wait_for_state(const State s) noexcept {
        if (get_state() < s) {
            std::unique_lock<std::mutex> lock{m};
            cv.wait(lock,
                    [this,s] () {return get_state() < s;}
            );
        } }
    }; \end{lstlisting}

        In the code fragment above, the lambda expression given to the
        standard library method {\tt wait} uses the method
        {\tt get\_state} to compare the state of the Thread instance to
        {\tt s}. It’s a guard against spurious wake-ups. {\tt wait} will
        always block until the condition variable is notified and the
        lambda expression returns {\tt true}.
        The behavior can be succinctly modeled in \textsc{Promela}, as
        shown below\footnote{
        The peculiar definition for {\tt LT}, the less-than relation,
        can be explained with the fact that {\sc Spin} treats the
        symbolic names in {\tt mtype} declarations in big-endian order,
        i.e. increasing from right to left. Note that there are no
        spurious wake-ups in \textsc{Promela}.}:
        \begin{lstlisting}
#define LT(x, y) x > y /* Sic */
inline wait_for_state(i, s) { !(LT(get_state(i), s)) } \end{lstlisting}

        Informally, the state ordering meant in {\tt wait\_for\_state}
        is the preorder of states implied by Figure
        \ref{fig:stateDiagram} when reading the arrows as ``less
        than''. Formally, the ordering is imposed by the enum
        definition for {\tt State}. In most situations these
        definitions are similar enough to not cause appreciable
        difference in behaviour. There is a special case, however: when
        a Thread in state {\tt PAUSED} is expected to proceed to
        the state {\tt RUNNING}. Since {\tt PAUSED} is formally greater
        than {\tt RUNNING}, an invocation
        {\tt wait\_for\_state(RUNNING)} on a paused Thread will
        immediately unblock, which is {\em incorrect}! This error was
        quickly found during the model checking with {\sc Spin}.

        C++ allows the programmer to specify integral constants for
        enumerators. This enables the use of bitmasks for conveniently
        expressing the exact set of target states where the waiting
        method needs to unblock. This detail was not modeled in
        \textsc{Promela}. Doing so would have required changing mtype
        for byte and losing symbolic names for states and commands in
        debug messages. Below is the reviewed C++ definition that
        enables the use of bit masks:
        \begin{lstlisting}
enum State {
    READY    = 1,
    STARTING = 2,
    /* ... */
    ABORTED  = 128 };
#define state_in(mask) (get_state() & mask) > 0 \end{lstlisting}

        The waiting function {\tt wait\_for\_state\_mask} utilizes
        bitmasks. It is the correct way to wait for a Thread to enter
        one of the states in the given bitmask. Below is the
        implementation:
        \begin{lstlisting}
void Thread::wait_for_state_mask(const uint8_t mask)
noexcept {
    if (!state_in(mask)) {
        std::unique_lock<std::mutex> lock{m};
        cv.wait(lock,
                [this,s] () {return state_in(mask);}
        ); }
}\end{lstlisting}

        This method can wait for a Thread in state {\tt PAUSED} to
        continue:
        \begin{lstlisting}
wait_for_state_mask( RUNNING  | STOPPING | STOPPED |
                     ABORTING | ABORTED );\end{lstlisting}
        As mentioned, the \textsc{Promela} definition for states was
        not changed, so instead of taking a bitmask, a hard-coded
        inline waiting block was added for each of the the five
        bitmasks that were needed. For instance, the C++ method
        invocation above was hard-coded into \textsc{Promela} as
        follows:
\lstset{
    classoffset=1,
    otherkeywords={wait_for_RESUME_mask},
    morekeywords={wait_for_RESUME_mask},
    keywordstyle=\color{black}}
        \begin{lstlisting}
inline wait_for_RESUME_mask(i) {
    get_state(i) == RUNNING || get_state(i) == STOPPING ||
    get_state(i) == STOPPED || get_state(i) == ABORTING ||
    get_state(i) == ABORTED; } \end{lstlisting}

    \section{DIVINE Models and Results}
        \label{divineModels}
        The interplay of state and command setters, state transitions,
        and waiting methods is non-trivial. Not all of the subtleties
        involved were exposed by the \textsc{Promela} model, partly
        because of the semantical difference between the
        \textsc{Promela} and C++ languages, but also because some of
        the details were left out from the model. This provided
        motivation for implementation-level model checking.

        We created the class {\tt DummyWorker} for modeling workers in
        \textsc{Divine}. It inherits the appropriate ADAPRO class and
        behaves like the \textsc{Promela} worker model. The method
        {\tt prepare} lets \textsc{Divine} choose an integer from
        $[0..3]$ non-deterministically, expressed as
        {\tt \_\_vm\_choose(4)} for selecting between the asynchronous
        trigger methods {\tt pause\_async} and {\tt stop\_async},
        throwing an exception, and doing nothing. Similarly, the methods
        {\tt execute} and {\tt finish} use the non-deterministic
        choice feature for implementing the behaviour described in
        Section \ref{verification}. The following C++ code example
        illustrates the implementation:

\lstset{
    classoffset=1,
    otherkeywords={atomic,mutex,condition_variable,unique_lock,
        lock_guard,uint8_t,State,Command,Thread,byte,failure,
        system_error,exception,Dummy,Worker,runtime_error},
    morekeywords={atomic,mutex,condition_variable,unique_lock,
        lock_guard,uint8_t,State,Command,Thread,byte,failure,
        system_error,exception,Dummy,Worker,runtime_error},
    keywordstyle=\color{navy},
    }
    \begin{lstlisting}
class DummyWorker final : public Worker {
protected:
    virtual void prepare() {
        switch (__vm_choose(4)) {
            case 0: pause_async(); break;
            case 1: stop_async(); break;
            case 2: throw std::runtime_error{"Error"};
            default: break;
        }
    }
    virtual void execute() { /* Stop, throw, or skip */ }
    virtual void finish() { /* Throw or skip */ }
} \end{lstlisting}

        We created three models for \textsc{Divine} using this class;
        the small model with one {\tt DummyWorker}, the medium model
        with {\tt Supervisor} and one {\tt DummyWorker}, and the large
        model that runs a full ADAPRO Session, with its built-in
        Supervisor and two {\tt DummyWorker} instances, using a
        hard-coded configuration. All models also have a main thread,
        that creates the appropriate ADAPRO objects, waits until they
        have carried out their tasks, and exits the program.

    \subsection{Premature Destructor Call Defect}
        As mentioned above, the smallest {\sc Divine} model featured
        just the main thread and one worker Thread. In the model, the
        main thread starts the worker and calls
        {\tt wait\_for\_state\_mask(STOPPED | ABORTED)}. {\sc Divine}
        found an execution where the main thread was woken up when the
        worker was performing an earlier state transition. Before the
        main thread could proceed to check whether or not the worker
        had reached the desired state, a context switch was performed,
        letting the worker to proceed. Next, the worker would set its
        state to {\tt STOPPED}, but before it was fully stopped, the
        main thread was scheduled again for execution. At that point,
        the main thread observed that the worker had entered into the
        {\tt STOPPED} state and proceeded to exit the program.

        At this point, a known issue about destroying a Thread with its
        backend implementation still running was triggered. The worker’s
        destructor was called as a part of the automatic cleanup
        following the RAII principle. When the backend thread of the
        worker was scheduled again for execution, it tried to refer to
        the ADAPRO Thread object that didn’t exist anymore which then
        caused the program to exit abnormally (and seemed to crash
        DiOS as well, through its Pthreads implementation).
        The premature destructor call issue has been the cause of the
        most catastrophical and hard to debug problems in the entire
        ADAPRO framework.

        With the current design of ADAPRO Threads, this issue is not
        easy to fix. However, there exists a workaround which was found
        to be correct in all situations by {\sc Divine}. This
        workaround is to simply make the virtual destructor of the final
        inheriting class to call {\tt Thread::join} (which then calls
        {\tt std::thread::join} to join the background thread) before
        releasing any resources.

    \subsection{Non-Terminated String Defect}
        A classic low-level programming mistake was found with
        {\sc Divine} when verifying the unit tests for Thread. The
        {\tt abort} method of a Thread was designed to permit the user
        to trigger the state transition through {\tt ABORTING} to
        {\tt ABORTED} with an error message given as an
        {\tt std::string} instance. The method dynamically allocated a
        C-style string, i.e. an array of ASCII characters, and called
        {\tt std::strcpy} using the pointer and length provided by the
        {\tt std::string} instance. The length reported was one less
        than needed for the character array, because a C-style string
        must end with an additional non-printable null character. The
        resulting C-style string would contain all the same characters
        as the {\tt std::string} object, but wouldn’t be
        null-terminated.

        Depending on the contents of the heap memory, reading a
        non-terminated string might succeed without problems, produce
        some extra garbage data past the intended ending of the string,
        or even cause a segmentation violation. There was also a lesser
        problem, namely that in some situations the C-style string was
        never deleted, causing a memory leak. Even Valgrind was not
        able to detect this leak, because it never actualised during
        debugging.

        Both the off-by-one error with the string length and the memory
        leak were quickly detected by {\sc Divine}, thanks to its
        sophisticated mechanisms that keep track on the objects
        allocated from heap and the pointers referring to them. Unit
        tests couldn't reliably detect this problem, as the next byte
        in heap usually happened to be zero. Even though this problem
        would have been easy to fix, the authors decided to abandon the
        {\tt abort} method altogether as an unnecessary complication to
        the overall design.

        This defect had been present in the framework for a very long
        time and wasn’t even detected during the hundreds of hours of
        simulations performed with ADAPOS applications\cite{Laang},
        built on top of ADAPRO, under maximum load, because the
        applications never encountered unhandled exceptions during the
        simulations.

    \subsection{Premature Command Defect}
        A model checker can only find errors reachable from the model
        under verification. A known issue with starting a Thread
        asynchronously, and then sending it a command before it has
        carried out its startup sequence, resulting in the command being
        ignored, was not detected by {\sc Divine}. The reason is that
        the models used the framework the way it was intended to be
        used, never exposing this particular scenario.

    \section{Experiences}
        In this section, we discuss the experiences we had with the
        tools, techniques, and languages used in this project.

    \subsection{Mapping C++ to \textsc{Promela}}

        The strength of \textsc{Promela} is its language level
        atomicity and synchronization concepts. This allows
        synchronization patterns to be expressed more concisely than in
        C++. The difference in the amount of code can be observed in
        the code examples above.

        There are also programming patterns that cannot be nicely
        translated from C++ to \textsc{Promela}. For example, in C++
        Thread is a class. The state of a Thread is represented as a
        (private) field (i.e. member in C++ parlance) within the class,
        which can be safely and conveniently accessed in other contexts
        through accessors. This sort of encapsulation is not possible
        in {\sc Promela}, and Threads need to be modeled using other
        means.

        We modeled Threads with processes and global arrays. In general,
        the field $f$ is represented by an array $a_f$, with $a_f[i]$
        storing the value of $f$ of the $i$th Thread. Such encoding
        clutters the model and moreover makes the model itself more
        error prone.

        Not having methods, classes, or inheritance at disposal with
        \textsc{Promela} added extra challenges, because these
        programming language features had to be substituted with CPP
        macros and \textsc{Promela} inline blocks. This complicated
        debugging the model in this project. While the resulting model
        in this project is still of manageable size, for a larger
        project this mismatch may eventually lead to a maintainability
        problem with hand-written models.

        The eight states and five commands of a Thread were modeled
        with {\tt mtype} declarations in \textsc{Promela}. As
        {\tt mtype} declarations share the same domain (of one byte
        in size), having two of them opened the possibility of
        erroneously assigning a state value to a command variable and
        vice versa. This risk had to be taken, since representing the
        two enumerated types with two distinct byte values might have
        grown the state space of the model. As mentioned, bytes cannot
        be pretty-printed which makes debugging less convenient than
        with {\tt mtype}s. On the other hand, bytes can be bitmasked,
        which is convenient as we have seen. Bitmasks of states and
        commands are utilized in the C++ implementation of the
        framework.

    \subsection{Verifying C++ With {\sc Divine}}
        This project initially used an older version of {\sc Divine}
        4, but had to change to the statically linked binary version
        4.1.20+2018.12.17 because of a bug in the DiOS operating system
        of an older {\sc Divine} version. The properties verified with
        {\sc Divine} included the absence of deadlocks, crashes, memory
        acces violations, and memory leaks.

        {\sc Divine} 4 is composed of a specialized version of the LLVM
        Clang compiler, a Phtreads implementation, a C++ standard
        library implementation, a minimalistic operating system,
        DiOS, the DiVM virtual machine and a model checker
        \cite{BBK+17, Rockai2018}. The {\sc Divine} model checker only
        observes the state transitions of DiVM. This design allows the
        {\sc Divine} model checker to also find errors in the operating
        system and libraries, in addition to verifying the application
        proper. A handful of defects in these software layers were
        found and reported to the {\sc Divine} developers during this
        project.

        As it turns out, model checking a non-trivial application with
        {\sc Divine} might not be as simple as just compiling and
        verifying the model, even if {\sc Divine} accepts C++ as an
        input language. It was found that C++ source code files
        must be compiled in a single invocation of the
        {\tt divine cc} command. Libraries, other than the standard
        library or Pthreads can’t be used, unless included in the same
        build with the user program.

        The C++ codebase of ADAPRO had to
        be modified with macros, undocumented internal {\sc Divine}
        compiler attributes (for optimisation) and DiVM hypercalls
        for getting it to work with {\sc Divine}. In general, the
        extent of modifications required probably depends heavily on
        the code under verification. We needed to put effort into
        the conversion, largely because some parts of the Linux and
        POSIX APIs were missing from the {\sc Divine} library layer.
        Thus, these parts of the code needed to be guarded by macros.
        The missing functionality included access to system clock,
        setting the {\tt nice} value of the running process, and
        setting scheduling options such as thread affinities. We also
        had to disable many non-critical parts of the code, e.g.
        logging, a watchdog mechanism, and change most of constans
        into functions, to keep the state space size feasible.

    \subsection{State Space Explosion}
        The hardest challenge in this project was the notorious state
        space explosion problem.  As discussed in the beginning of
        Section \ref{findings}, we could only fully verify the
        {\tt Promela} model with two workers.

        For making verification with {\sc Divine} possible, the
        implementation of ADAPRO needed several iterations of
        refactoring and optimisation to keep the state space size of
        even the small model manageable. One particular header file
        ({\tt headers/data/Parameters.hpp}) had many global
        constant variables. In C++ global variables have internal
        linkage, which means that every translation unit gets their own
        unique copies of these variables, stored in different memory
        locations in the computer executing the compiled machine code.
        Among other optimizations, replacing these global variables
        with functions returning literals on demand, provided around
        60\% reduction in the size of DiVM states during verification
        runs.

        The number of states was another aspect of the problem. For the
        small model, running {\tt divine check} took around 5,000 DiVM
        states. When the medium-size model was finally checked
        successfully, the verification took almost 500,000 states. By
        the time of writing this paper, the large model and an example
        application involving disk I/O through a virtual filesystem
        image still remain to be checked.

        Disabling the logger facility of ADAPRO and all other standard
        output and error stream operations also proved to make a huge
        difference in the size of the state space. Before disabling
        output operations and watchdog mechanisms, a verification run
        for the medium-size model on a virtual machine, in a cloud
        provided by SURFsara, had been running for more than 17 hours
        of wall clock time, consuming over 37 gigabytes, with no end in
        sight. After the optimisations, the medium-size model can now
        be checked in twenty minutes on a laptop, with the peak memory
        usage being in the order of seven gigabytes. Even after all
        optimisations, a virtual machine provided
        by SURFsara couldn’t finish the verification run for the large
        model, given 124 GB of RAM dedicated to \textsc{Divine}.

    \subsection{Scalability}
        When formulating the Theory of ADAPRO, we found out that
        {\sc Spin} was not able to digest the properties (1) -- (8) in
        a single formula. We first tried to do this and {\sc Spin}
        encountered a buffer overflow when parsing the property. On
        second thought, having one huge property would probably have
        been prohibitively expensive to verify anyway.

        We found out, that even though {\sc Divine} supports
        multi-threaded verification, the performance doesn’t scale for
        an arbitrary number of threads. Especially on the virtual
        machines provided by SURFsara, the scaling proved to be
        unsatisfactory. As mentioned before, verifying the {\sc Promela}
        model takes only a couple of minutes, so multi-threaded
        verification with {\sc Spin} was not needed.

        It seems that, for the ADAPRO models, two
        verification threads yield the best overall throughput in terms
        of DiVM states and instructions explored per second. The higher
        the number of threads, the higher the ratio of system time to
        user time. Thus, beyond the saturation point, increasing the
        number of threads seems to only increase the CPU time
        consumption and heat production. In fact, verification speed
        seems to slowly decline as a function of thread count.
        {\sc Divine} developers were not aware of such performance
        issues, so our models might have been anomalous, or perhaps the
        virtual machine or the guest system was configured suboptimally.

    \subsection{Thoughts on the Learning Curves}
        All in all, the experiences with the {\sc Spin} and
        {\sc Divine} model checkers have been positive and encouraging
        in this project. Even though the \textsc{Promela} language was
        quite different from C++, it felt easy to learn. Likewise,
        after some initial learning effort, using {\sc Divine} to check
        almost any standard C++ code proved to be easy. Even with their
        own challenges, using both of the model checkers was easy
        compared to tactic-based interactive theorem proving with HOL
        on Poly/ML or the use of propositions-as-types interpretation
        of intuitionistic type theory in Agda.

        Of course, programmers need to get at least some degree of
        familiarity with the basics of enumerative model checking, in
        order to understand how {\sc Spin} and {\sc Divine} are best
        exploited. We estimate that it'd take a few weeks of training
        to acquaint developers with a few years of experience in
        programming and some mathematical maturity, into the use of
        these model checkers. 
        Perhaps the effort could be compared with learning
        a new programming language.

        We found the output of {\sc Spin} model checker harder to
        understand than the reports generated by static analysis,
        testing, and code coverage tools, or the {\sc Divine} model
        checker. Furthermore, the semantical gap between
        \textsc{Promela} and C++, especially in terms of their memory
        models and atomicity, is considerable. This might introduce
        inaccuracies to hand-made translations between C++ and
        {\sc Promela}. Luckily, this was not a problem since we also
        performed model checking on the very C++ source code with
        {\sc Divine}.

    \subsection{Impact on Workflow}
        Installing the {\sc Spin} and {\sc Divine} model checkers
        was easy and straightforward. The Debian GNU/Linux operating
        system has {\sc Spin} included in its standard package
        repository, so the installation is just a matter of running
        {\tt apt-get install spin}. Using the precompiled binary version
        of {\sc Divine} is just as easy. We considered it sufficient
        for this project to run {\sc Divine} by invoking a custom BASH
        runner script.

        Thanks to their simple command-line interfaces, integrating
        {\sc Spin} and {\sc Divine} into the workflow of the software
        project was fast and easy (after the initial work spent in
        learning the correct flags and arguments). These tools don’t
        require maintaining complicated configuration files, but their
        use can be fully parameterized and automated with regular shell
        or makefile scripts, for example. Setting up the integration
        takes maybe even less time than with an average software tool.

        After integrating the model checkers into the project workflow,
        they can be used without hassle on daily basis, just like any
        other analysis or testing tools. Model checking unit tests with
        {\sc Divine} was found especially useful, because that way,
        existing test cases can be explored exhaustively. As noted
        above, sheer code coverage is not enough to ensure
        exhaustiveness of tests. Developing models and implementation
        in parallel seems to be an efficient analogue of test-driven
        development.

    \section{Conclusion}
        The LTL properties listed in Section \ref{subsec.theory},
        should, from now on, be considered an integral part of the
        specification of ADAPRO. Model checking was found to be a
        valuable addition to software developer's toolbox, with a good
        return of investment value (which already seems to be the
        consensus in the literature). It seems that different model
        checkers can find different kinds of issues, on different
        levels of abstraction (e.g. design/implementation). Many of the
        issues found during this project require rather specific
        circumstances to occur, making them nearly impossible to test,
        but when they do occur, they may trigger chain reactions of dire
        consequence.

        We believe that easy integration to the everyday workflow is
        important for any software tool to become successful in the
        industry. In terms of automation possibilities, it seems that
        we’re already getting there with model checkers. However, APIs,
        output formatting, and documentation still may to need extra
        polishing, not to mention IDE support. We believe that these
        points need to be addressed, before formal methods can attract
        the attention of mainstream programmers.

        The benefits of model checking far outweigh the slightly rough
        edges of the model checkers. The shortcomings could be
        mitigated, with more resources invested in the development of
        formal verification tools. The sheer number of rather simple
        defects that were found in supposedly stable code, only after
        deploying model checkers, raises a question: \emph{Can
        software, that has not been formally verified, be trusted?} We
        doubt it.

    \subsection{Future Prospects}
        The effect on fairness imposed by user Threads getting
        exclusive access to CPU cores through the use of the Pthreads
        API, has not been explored. As mentioned before, configuration
        access and Interaction with external systems using signals, the
        DIM protocol, and the Linux systemd API was not modeled.
        Investigating these aspects might prove useful, albeit
        challenging.

        Even though the {\sc Promela} model was successfully verified
        with two workers, there's no hard mathematical argument showing
        that two workers suffice. Decreasing the granularity of the
        {\sc Promela} model might allow more workers to be simulated.
        Using a swarm-based approach\cite{Holzmann2008} might turn out
        to be a useful bug-hunting technique with larger models for
        both {\sc Spin} and {\sc Divine}.

        The {\sc Divine} models still deserve more attention.
        Firstly, measuring the effects of different kinds of code
        optimizations on the number and size of {\sc Divine} states
        would be an interesting topic for a quantitative study.
        Secondly, using monitors to perform LTL model checking on ADAPRO
        was not yet attempted, because our {\sc Divine} models were
        too large even without LTL properties. Thirdly,
        verifying ADAPRO under a weak memory model was also not
        attempted, since {\sc Divine} does not yet support such a
        feature.

\subsection*{Acknowledgements}
        We thank Prof. Keijo Heljanko from the Dept. of
        Computer Science, University of Helsinki, for consultation. For
        guidance on using {\sc Divine}, we thank dr. Petr Ročkai
        from the Faculty of Informatics, Masaryk University, Brno. For
        the work done on ADAPRO, we thank Peter Chochula, Peter Bond,
        and the rest of our colleagues in ALICE DCS central team at
        CERN. We thank Harri Hirvonsalo from CSC -- IT
        Center for Science Ltd. and Jarkko Savela from University of
        Helsinki for providing feedback. Some of the
        experiments were carried out on the Dutch national
        e-infrastructure with the support of SURF Cooperative, for
        which we are grateful. The main author is grateful for
        receiving the ACM SIGSOFT CAPS grant.


    \bibliographystyle{ACM-Reference-Format}
    \bibliography{paper}
\end{document}